\documentclass{article}
\usepackage{spconf,amsmath,graphicx}
\usepackage{multirow}
\usepackage{url}
\usepackage{makecell}

\title{IMPROVED SELF-SUPERVISED MULTILINGUAL SPEECH REPRESENTATION LEARNING COMBINED WITH AUXILIARY LANGUAGE INFORMATION}
\name{Fenglin Ding $^{1, 2}$, Genshun Wan$^{1, 2, 3}$,Pengcheng Li$^{1, 2}$,  Jia Pan$^{1, 2}$, Cong Liu$^{1, 2}$}
\address{$^1$Shanghai AI Laboratory, China \\
$^2$iFLYTEK Research, iFLYTEK Co. Ltd., China \\
$^3$University of Science and Technology of China, China \\
\small{\texttt{\{flding2, gswan, pcli2, jiapan,congliu2\}@iflytek.com}}}

\begin{document}
\ninept
\maketitle
\begin{abstract}
Multilingual end-to-end models have shown great improvement over monolingual systems. With the development of pre-training methods on speech, self-supervised multilingual speech representation learning like XLSR has shown success in improving the performance of multilingual automatic speech recognition (ASR). However, similar to the supervised learning, multilingual pre-training may also suffer from language interference and   further affect the application of  multilingual system. In this paper, we introduce several techniques for improving self-supervised multilingual pre-training by leveraging auxiliary language information, including the language adversarial training, language embedding  and language adaptive training  during the  pre-training stage. We conduct experiments on a multilingual ASR task consisting of 16 languages. Our experimental results demonstrate 14.3\% relative gain over the standard XLSR model, and 19.8\% relative gain over the no pre-training multilingual model.

\end{abstract}
\section{Introduction}
\label{sec:intro}
The performance of automatic speech recognition (ASR) has progressed substantially in the recent past~\cite{1-synnaeve2019end,3-gulati2020conformer,4-chan2021speechstew}. However, ASR has suffered from the reliance on human annotations, which is typically impractical for most languages around the world. For multilingual business, there is considerable interest and importance to support multiple languages in a single ASR system. To take advantage of the commonality among different languages, multilingual models have been studied over the last two decades and have outperformed monolingual models for ASR in low-resource languages ~\cite{6-kannan2019large,7-pratap2020massively}.

Despite such promising potential of multilingual models, most approaches inevitably require abundant training data because it combines acoustic models, language models and alignment into the same model. Therefore, there has been a lot of interest in how to better use unlabeled speech data since the availability of unlabeled data may be virtually unlimited~\cite{9-liu2019adversarial,11-hsu2020semi} . As a mainstream study direction, self-supervised Learning (SSL) typically utilizes speech pre-training to learn better representations. Based on a pre-trained model, the recognition system can achieve comparable performance to supervised models trained on quantities of labeled data, while only requiring fine-tuning on a small amount of labeled data ~\cite{15-baevski2020wav2vec,21-chen2022wavlm,22-baevski2022data2vec}.

Self-supervised learning has been  applied to multilingual modeling in recently, namely cross-lingual speech representation learning (XLSR)~\cite{23-conneau2020unsupervised}. And other studies also show that multilingual pre-training outperforms monolingual pre-training in low resource languages~\cite{24-babu2021xls,25-conneau2020unsupervised}. However, according to the studies in ~\cite{24-babu2021xls}, there exists the capacity dilution and interference problem in multilingual pre-training. Specifically, low resource languages benefit from joint training with similar languages, while high resource languages often suffer from the negative transfer problem, resulting in inferior performance. Such performance degeneration becomes more significant as the model expands to more languages ~\cite{7-pratap2020massively} or more training data ~\cite{26-li2021scaling}, which obstacles the application of pre-trained models on multilingual tasks.

Motivated by that, we propose several methods to alleviate language interference to improve the performance of multilingual pre-training by leveraging auxiliary language information. These approaches include the introduction of language adversarial training, language embedding  and language adaptive training  during the  pre-training stage, so as to balance the learning of shared representations and language specific representations. At inference time, we assume the language is either specified in the user's preferences, or determined automatically from a language identification system. We analyze the trade-offs of different methods with regards to quality gains and number of additional parameters, and further explore the key factors that affect the effectiveness of each method. We conduct experiments in an ASR task consisting of 16 languages. Our experimental results demonstrate $\sim$4\% to $\sim$14\% relative gain over the standard XLSR model, and $\sim$10\% to $\sim$20\% relative gain over the baseline multilingual model.

\section{MODELING}
\label{sec:related}

Self-supervised approaches such as wav2vec 2.0  ~\cite{15-baevski2020wav2vec} have been extended to the multilingual setting. The recent XLSR or XLS-R focus on the cross-lingual setting by learning representations on unlabeled data that generalize across languages.

XLSR builds on the pre-training approach of wav2vec 2.0 which jointly learns contextualized speech representations as well as a discrete vocabulary of latent speech representations. The backbone of XLSR includes a feature encoder, a context representations encoder and a quantization module. During pre-training, the feature encoder takes the speech frames as input and generate higher dimensional features $\boldsymbol{Z} = [\boldsymbol{z}_1,\boldsymbol{z}_2,\dots,\boldsymbol{z}_T]$. Then some frames of $\boldsymbol{Z}$ are randomly selected to be masked. The corrupted features $\boldsymbol{\tilde{Z}}$ are fed into transformer based context network and then transformed into contextual representations $\boldsymbol{C} = [\boldsymbol{c}_1,\boldsymbol{c}_2,\dots,\boldsymbol{c}_T]$ at the top layer~\cite{31-vaswani2017attention}. The model is trained by solving a contrastive task over masked feature encoder outputs. To provide the training targets, unmasked feature encoder representations Z are discretized to $[\boldsymbol{q}_1,\boldsymbol{q}_2,\dots,\boldsymbol{q}_T]$ with a quantization module. The quantization module uses a Gumbel softmax ~\cite{32-jang2016categorical} to choose entries form the codebooks and the chosen entries are concatenated to obtain $\boldsymbol{q}$.  The goal of the contrast task is to distinguish a true encoded sample qt among distractors $\boldsymbol{Q}_t$ that are sampled uniformly from other masked steps of the same sequence:
\begin{equation}
	L=-log\frac{exp(sim(\boldsymbol{c}_t,\boldsymbol{q}_t)/\tau)}{\sum_{\boldsymbol{q}^{'}\in \boldsymbol{Q}_t}exp(sim(\boldsymbol{c}_t,\boldsymbol{q}^{'})/\tau)}
\end{equation}
Where $sim(a, b)$ computes the cosine similarity and $\tau = 0.1$ scales the logit. The  objective  is  augmented  by  a  codebook diversity  penalty to encourage the  model  to use  all  codebook entries.

We train the standard XLSR as a pre-training baseline in our work. Furthermore, we replace Transformer with Conformer ~\cite{33-gulati2020conformer} as the context representations encoder, which has presented higher effectiveness in ASR task. The major component of Conformer is a stack of conformer blocks, each of which is a series of multi-headed self-attention, depth-wise convolution and feed-forward layers. On the other hand, to improve computational efficiency, we use filter-bank features as input of network instead of raw waveforms. The feature encoder that acts as a convolutional sub-sampling block is replaced with two 2D-convolution layers, both with strides (2, 2), resulting in a 4x reduction in the acoustic sequence length.

\section{PRE-TRAINING WITH LANGUAGE INFORMATION}
\label{sec:proposed}
In this section, we present the different approaches to improve self-supervised multilingual pre-training. Based on the standard XLSR model, we introduce language information into multilingual representation learning in different designs, to alleviate the negative impact of language interference.

\subsection{Language adversarial pre-training}
In order to guide the model to learn language-independent representations that are more amenable to adaptation, we propose to combine language adversarial training ~\cite{27-yi2018language} with multilingual pre-training, called language adversarial pre-training.

In language adversarial pre-training, a language discriminator is used to recognize the language label using the shared features of the multilingual pre-training model. An additional language label is given for each training sample $\left \{ h_i^{m},m \right \}$ , where $h_i^{m}$ denotes the output of a hidden layer,  $m \in \left \{ 1,\dots,M, \right \} $ denotes the language label for each frame, and M is the number of languages in training sets. The language discriminator loss function $L_{adv}$ is defined as:
\begin{equation}
	L_{adv}=-\sum_{m=1}^{M} \sum_{i=1}^{N_m} logP(m|h_i^{m};\theta^{s},\theta^{d})
\end{equation}
Where $N_m$ denotes the number of training samples of language $m$. $\theta^{s}$ denotes the parameters of the shared layers, $\theta^{d}$ denotes the parameters of the sub-network of the language discriminator.

Meanwhile, the gradient reversal layer (GRL) ~\cite{34-ganin2016domain}  is introduced to ensure the feature distributions over all the languages are as indistinguishable as possible for the language discriminator. Thus the shared layers can learn language-invariant features. At the feed-forward stage, the GRL acts as an identity transformation. During the back-propagation, however, the GRL takes the gradient from the subsequent level and changes its sign, i.e., multiplying by -1. The GRL has no parameters associated with it.

Thus, using the multi task learning method, the language adversarial pre-training is to simultaneously optimize contrastive loss and language discriminator loss:
\begin{equation}
	L=L_c-\lambda L_{adv}
\end{equation}
Where $L_c$ is the contrastive loss described in Eq.(1), and $\lambda$ controls the trade-off between the two loss.

\subsection{Language embedding involved pre-training}
Then we investigate the architectural extension of multilingual pre-training by feeding a vector to represent the language ~\cite{28-toshniwal2018multilingual}. It is assumed that a universal model can use a language vector to become an ``expert'' on each individual language instead of learning a representation tailored to languages with more data. Various methods of using a language vector have been previously described and directly compared in E2E multilingual ~\cite{35-toshniwal2018multilingual} and multi-dialect ~\cite{36-li2018multi} models.

The language itself can be represented in a one-hot vector or an embedding vector. In addition, given such a language vector, there are many different position in the model where it can be inserted. In this work, language embedding shows its effectiveness when involved in multilingual pre-training. Moreover, our experiments have confirmed earlier observations that simply adding it as language-specific bias terms to the low-level features is sufficient. Thus, given a training sample with language label  $\left \{ h_i^{m},m \right \}$, the multilingual pre-training representation $h_i^{m}$ is enhanced by language embedding as:
\begin{equation}
	\tilde{h}_i^{m}=h_i^{m}+embedding(m)
\end{equation}
Where the embedding of language $m$ is trainable parameters.

Furthermore, in order to hold the learned language embedding as different as possible and provide more effective language specific information, the language orthogonal constraint loss is defined as:
\begin{equation}
	L_{o}=-\sum_{i=1}^{m} \sum_{j=1}^{m} ||E_iE_j^{\top}-[i==j]||^{2}
\end{equation}
Where $E_i$ and $E_j$ are the embedding of language $i$ and $j$, respectively. Combined with contrastive loss, the final multilingual pre-training loss is defined as:
\begin{equation}
	L=L_c+\alpha L_{o}
\end{equation}
Where $L_c$ is the contrastive loss described in Eq.(1), and $\alpha$ is a penalty coefficient of the constraint.

\subsection{Language specific adaptive pre-training}
\label{subsec:teacher}
Intuitively, multilingual model needs to learn the common representation between different languages to improve the performance of low resource languages from transfer learning, while retaining the language specialization to avoid the negative impact of unrelated languages. Therefore, researchers opt to use specific network components for each language. Adapters ~\cite{29-le2021lightweight} and adaptive weights ~\cite{30-pham2021efficient} have got more attention in supervised multilingual models due to being computationally manageable and achieve promising results. Here we extend them to self-supervised multilingual pre-training, becoming language specific adaptive pre-training.

Adapters are implemented by a small multi-layer perceptrons (MLP), in which one hidden layer acts as a down-sampler for parameter efficiency. This MLP is serialized at the end of each layer in  Transformer to help the network changes the feature distribution based on languages.

However, the language adapters added to the Transformer are essentially feed-forward neural network layers similar to the counterpart already in the shared Transformer body. A scenario with 20 languages consequently generates hundreds of these layers accounting for a large amount of parameters to be optimized. To solve this problem, adaptive weights was proposed by using a factorization scheme that is both effective and highly scalable for multilingual system.

The main idea of adaptive weights is that each matrix multiplication $Y=W^{\top}X$ in the multilingual model can be decomposed into a function of shared weights $W_S$ and additional language dependent weights $W_{ML}$ and $W_{BL}$:
\begin{equation}
	Y=(W_S\cdot W_{ML}+W_{BL})^{\top} X
\end{equation}

In order to encourage the model to share parameters as well as keeping the parameters efficient, the adaptive weights are factorized by using the form of 1-rank matrices which can be compactly represented as a dot-product between two vectors. One drawback in this method is the lacking representational power of Rank-1 matrices. One solution is to modify the factorization into using $k$ vectors per language so that there are k independent weight factors followed by a summation, which increases the rank of additional weight matrices:
\begin{equation}
	\bar{W} =\sum_{i}^{k} r_is_i^{\top }
\end{equation}

\section{Experiments}
\subsection{Datasets}
We conduct our experiments on a inner multilingual speech recognition task, which consist of 16 languages from different countries and regions. Train and test set sizes vary due to availability of transcribed data and are shown in \tablename~\ref{tab:tab1}. For all of these languages, the training data is anonymized and transcribed by humans. Transcription of each language is cleaned and only the scripts of the language are included. The test sets are recorded independently of the training set and transcribed manually. Therefore, in some languages, there may be serious mismatches between the training set and the test set.

\begin{table}
\centering
\caption{Number of hours in train sets and utterances in test sets.}
\label{tab:tab1}
\begin{tabular}{c|cc|c|cc}
\hline
Lang. & train/h & test/utt. & Lang. & train/h & test/utt.  \\
\hline
ml-in & 328     & 1k        & ps-ar & 295     & 1k         \\
ne-np & 397     & 1k        & he-il & 477     & 1k         \\
hy-am & 342     & 2k        & af-za & 313     & 1k         \\
ka-ge & 281     & 1k        & az-az & 246     & 2k         \\
mr-in & 269     & 1k        & da-dk & 320     & 1k         \\
te-in & 206     & 1k        & fi-fi & 287     & 1k         \\
am-et & 266     & 1k        & lo-la & 434     & 1k         \\
tg-ke & 364     & 2k        & my-mm & 362     & 2k         \\
\hline
\end{tabular}
\end{table}

\subsection{Model Setups}
We train a XLSR model as the pre-training baseline. The whole encoder consists of a feature encoder and a context representation encoder. Feature encoder has 2 convolutional layers with filter size (3, 3), strides (2, 2). Two layers have 128 and 32 channels respectively. Representation encoder consists of 16 Conformer blocks, each with encoder embedding 512, hidden dimension 2048, 8 attention heads and convolution kernel size 15. The final projection layer dimension is 768. In total, the whole encoder including feature encoder has about 100 million trainable parameters. We also train a multilingual ASR model as the supervised baseline. The encoder is consistent with the pre-training model. For the decoder, simple yet effective, we use a Transformer decoder. Decoder consists of 6 blocks, each with decoder embedding 768, hidden dimension 3072, 12 attention heads. The input features are 40-dimensional filter-bank acoustic features computed every 10ms over a 25ms window.

In language adversarial pre-training, the language discriminator consists of two layers of perceptrons, each with 512 units, and a classification layer with 16 units. For language embedding involved pre-training, the dimension of language embedding is consistent with its added feature dimension. During language specific adaptive pre-training, the structure of adapter module follows [29], and the projection dimension is set to 256. While the number of $k$ in adaptive weights is explored in the following experiments.

\subsection{Training Details}

\textbf{Pre-training}. We train all the self-supervised models on 32 A100 GPUs for 400k steps.  We upsample the data from low resource languages with $\alpha$ = 0.5 and for high resource languages we use natural sampling probability with $\alpha$ = 1.0. We use dropout 0.1 in the conformer and at the output of the feature encoder. Layers are dropped at a rate of 0.2. We optimize with Adam ~\cite{38-kingma2014adam}, warming up the learning rate for the first 8000 updates to a peak of 0.0005, and then linearly decay it. For masking, we sample p = 0.065 of all time-steps to be starting indices and mask the subsequent 10 time-steps.

\textbf{Fine-tuning}. During fine-tuning, we add the same decoder as the supervised multilingual model to the pre-training model. We also insert a linear layer between the pre-trained model and the decoder layer as the projection block. We fine-tune the model for 15 epochs. The learning rate warms up linearly from 0 to the peak learning rate 0.0007 for the first 8000 training steps, decays to the 10-th epoch and then is halved for each epoch from the 11-th epoch. During fine-tuning, the parameters of feature encoder are fixed.

\subsection{Results and Discussion}
The performance of the different pre-trained modules are fully presented in \tablename~\ref{tab:tab2}. Standard pre-training model, XLSR, outperforms the multilingual model without pre-training, achieving 6.5\% relatively improvement. It can be seen that with the increase of the number of languages and supervised data in the fine-tuning stage, the general pre-training model has not been improved as much as reported in the paper ~\cite{23-conneau2020unsupervised}, even performance of some languages has obviously deteriorated, which confirms the negative impact of language interference in the pre-training process. Furthermore, when multilingual pre-training is assisted by several pre-training strategies that we propose to introduce language information, it exceeds the general pre-training model.

Among the improved multilingual pre-training, the performance of language adversarial pre-training and language embedding involved pre-training is similar, with 4.3\% and 3.6\% improvement respectively compared with the standard pre-training model, and 10.5\% and 9.7\% improvement respectively compared with the supervised multilingual model. The language specific adaptive pre-training with adapter or adaptive weights, performs even better. Compared with the XLSR model, they achieve 11.3\% and 14.3\% improvement respectively, while compared with the supervised multilingual baseline, they achieve 17.0\% and 19.8\% improvement respectively. Finally, with regards to performance gains and number of additional parameters introduced, pre-training with language specific adaptive weight significantly improves the multilingual pre-training with few additional parameters.

\begin{table}
\centering
\caption{WER(\%) on the test set of 16 languages. Models include conformer (C), standard pre-training (XLSR), language adversarial pre-training (LA), language embedding involved pre-training (LE), pre-training with language specific adapter (LSA), and pre-training with language specific adaptive weight (LSAW).}
\label{tab:tab2}
\begin{tabular}{c|cccccc}
\hline
Lang.                                                   & CF             & XLSR  & LA             & LE             & LSA            & LSAW            \\
\hline
\begin{tabular}[c]{@{}l@{}}Params \\Inc.(\%)\end{tabular} & -              & 0     & 0.2            & 0.01           & 4.9            & 0.5             \\
\hline
ml-in                                                   & 37.67          & 38.71 & 38.46          & 37.61          & \textbf{37.47} & 37.6            \\
ne-np                                                   & 45.34          & 46.77 & 44.86          & 45.92          & \textbf{42.75} & 42.83           \\
hy-am                                                   & 21.41          & 21    & 16.51          & 18.96          & 17.01          & \textbf{14.72}  \\
ka-ge                                                   & 7.94           & 7.67  & 7.48           & 7.41           & \textbf{7.40}  & 7.51            \\
mr-in                                                   & 14.31\textbf{} & 15.1  & 15.05          & 14.94          & 14.80          & \textbf{13.87}  \\
te-in                                                   & 18.16\textbf{} & 18.81 & 18.52          & 18.78          & 18.22          & \textbf{17.45}  \\
am-et                                                   & 20.44\textbf{} & 20.24 & 20.39          & 20.06          & 20.19          & \textbf{19.81}  \\
tg-ke                                                   & 18.44          & 19.15 & 18.79          & 18.56          & \textbf{18.14} & 18.48           \\
ps-ar                                                   & 22.73\textbf{} & 23.38 & 22.83          & 22.48          & 22.61          & \textbf{20.48}  \\
he-il                                                   & 21.17          & 24.88 & 23.72          & 24.85          & 20.87          & \textbf{20.71}  \\
af-za                                                   & 24.14          & 22.74 & \textbf{22.4}  & 22.69          & 22.16          & 22.42           \\
az-az                                                   & 36.59          & 23.28 & 18.86          & \textbf{18.75} & 24.54          & 20.78           \\
da-dk                                                   & 38.11          & 24.69 & \textbf{24.37} & 25.45          & 25.45          & 24.5            \\
fi-fi                                                   & 56.66          & 52.07 & 47.88          & 49.92          & 27.90          & \textbf{26.71}  \\
lo-la                                                   & 20.30          & 13.34 & 12.9           & 13.87          & 12.16          & \textbf{12.33}  \\
my-mm                                                   & 37.33          & 40.54 & 41.61          & 37.79          & 34.25          & \textbf{33.22}  \\
\hline
AVG.                                                    & 27.55          & 25.77 & 24.66          & 24.88          & 22.87          & \textbf{22.09}  \\
\hline
\end{tabular}
\end{table}

\subsection{Ablation Studies}
\textbf{Language adversarial pre-training:} Intuitively, the more language-specific information contained in the features fed into the language discriminator, the more easily the network can learn the language-invariant features under the effect of the gradient reversal layer (GRL). And language information is similar to speaker information, which is a long-term stable information, usually contained in the low layer of the network. Therefore, we explore the impact of the position of the language discriminator inserted into the network and the loss weight of language classification.

\begin{table}[t]
\centering
\setlength{\tabcolsep}{7mm}{
\caption{Comparison of different insertion positions and loss weights in language adversarial training.}
\label{tab:tab3}
\begin{tabular}{ccc}
\hline
Layer & Weight & AVG. WER        \\
\hline
16    & 0.1         & 25.98           \\
16    & 0.01        & 25.16           \\
16    & 0.005       & 25.38           \\
8     & 0.01        & 24.97           \\
4     & 0.01        & \textbf{24.66}  \\
0     & 0.01        & 24.83           \\
\hline
\end{tabular}
}
\end{table}

The results of different insertion positions and loss weights of discriminator are summarized in \tablename~\ref{tab:tab3}, Layer 0 means the input of the context representation encoder of the pre-training model. First of all, it can be seen that too large or too small weight is not conducive to the language adversarial training, and even has a negative impact on multilingual pre-training. In addition, the lower layer insertion position works best, which is consistent with the assumption. In this work, the best performance is achieved by sending the output of the 4-th conformer block to the language discriminator.

\textbf{Language embedding involved pre-training:}Based on the language adversarial pre-training, we investigate the influence of the position of language embedding inserted into network on effectiveness of pre-training. Moreover, the effect of proposed language orthogonal constraint under different penalty coefficients is also discussed.

As shown in \tablename~\ref{tab:tab4}, inserting language embedding into the input of representation encoder performs best, which confirms the existing research ~\cite{35-toshniwal2018multilingual,36-li2018multi} and our observation in Language adversarial pre-training. Additionally, further improvement can be achieved by adding orthogonal constraints, and the best penalty coefficient is 10. This shows that language orthogonal constraints can help model learning more distinguishable language embedding.

\begin{table}[t]
\centering
\setlength{\tabcolsep}{7mm}{
\caption{Comparison of different insertion positions and orthogonal constraint in language embedding involved pre-training.}
\label{tab:tab4}
\begin{tabular}{ccc}
\hline
Layer & orthogonal constraint & AVG. WER        \\
\hline
0     & 0                     & 25.36           \\
8     & 0                     & 25.65           \\
16    & 0                     & 25.87           \\
0     & 1                     & 25.03           \\
0     & 10                    & \textbf{24.88}  \\
0     & 50                    & 25.32           \\
\hline
\end{tabular}
}
\end{table}

\begin{table}
\centering
\setlength{\tabcolsep}{7mm}{
\caption{Comparison of different $k$ in language specific adaptive pre-training with adaptive weights.}
\label{tab:tab5}
\begin{tabular}{cc}
\hline
$k$ in adaptive
  weight & AVG. WER        \\
\hline
1                      & 24.06           \\
(1,8)                  & 23.47           \\
8                      & \textbf{22.09}  \\
(8,16)                 & 22.32           \\
16                     & 22.13           \\
\hline
\end{tabular}
}
\end{table}

\textbf{Language specific adaptive pre-training:} Based on the conclusions of our previous two investigations and the consideration of additional parameters, we only insert adapter or adaptive weight into the input of representation encoder in language specific adaptive pre-training. We explore the effect of number of $k$ in adaptive weights. The results are presented in \tablename~\ref{tab:tab5}, for the value in parentheses, the former represents the number of vectors used for scale matrices, and the latter represents the number of vectors used for bias matrices. It can be seen that when $k$ for scale matrices and bias matrices both are eight, adaptive weight can achieve the best performance.

\section{Conclusion}
In this paper, we propose several methods to alleviate language interference to improve the performance of multilingual pre-training by leveraging auxiliary language information. These approaches include the extension of language adversarial training, language embedding and language adaptive training to pre-training, so as to balance the learning of shared representations and language specific representations. We analyze the trade-offs of different methods with regards to quality gains and number of additional parameters introduced, and further explore the key factors that affect the effectiveness of each method. We conduct experiments in an ASR task consisting of 16 languages. Our experimental results demonstrate $\sim$4\% to $\sim$14\% relative gain over the standard XLSR model, and $\sim$10\% to $\sim$20\% relative gain over the baseline multilingual model.



\bibliographystyle{IEEEbib}
\bibliography{strings,refs}

\end{document}